\documentclass{article}

\usepackage{arxiv}

\usepackage[utf8]{inputenc} 
\usepackage[T1]{fontenc}    
\usepackage{hyperref}       
\usepackage{url}            
\usepackage{booktabs}       
\usepackage{amsfonts}       
\usepackage{nicefrac}       
\usepackage{microtype}      
\usepackage{lipsum}
\usepackage{graphicx}
\graphicspath{ {./images/} }
\usepackage{tikz}
    
\usetikzlibrary{calc}
\usepackage[framemethod=default]{mdframed}
 \usepackage{tabularx} 
 \usepackage{multirow}
 \usetikzlibrary{backgrounds,calc,shadings,shapes.arrows,shapes.symbols,shadows}

\tikzset{cvcv/.style={
     cloud, draw, aspect=2,color={black}
  }
}

\usepackage{pgfplots}
\usepackage{bchart}
\usepackage{pgfplotstable}
\pgfplotsset{compat=1.7}
\usepackage{amsmath}
\usetikzlibrary{arrows}
\usepackage{pgf}
\usepackage{tikz} 
\usetikzlibrary{shapes,arrows,automata}
\usetikzlibrary{shapes.geometric,backgrounds,calc}
\tikzset{
  basic box/.style = {
    shape = rectangle,
    align = center,
    draw  = #1,
    fill  = #1!25,
    rounded corners},
  header node/.style = {
    Minimum Width = header nodes,
    font          = \strut\Large\ttfamily,
    text depth    = +0pt,
    fill          = white,
    draw},
  header/.style = {%
    inner ysep = +1.5em,
    append after command = {
      \pgfextra{\let\TikZlastnode\tikzlastnode}
      node [header node] (header-\TikZlastnode) at (\TikZlastnode.north) {#1}
      node [span = (\TikZlastnode)(header-\TikZlastnode)]
        at (fit bounding box) (h-\TikZlastnode) {}
    }
  },
  hv/.style = {to path = {-|(\tikztotarget)\tikztonodes}},
  vh/.style = {to path = {|-(\tikztotarget)\tikztonodes}},
  fat blue line/.style = {ultra thick, blue}
}
\usepackage{listings}
\usetikzlibrary{positioning}
\title{Toward Efficient Web Publishing with Provenance of Information Using Trusty URIs: Applying the proposed model with the Quran}

\author{ \href{https://orcid.org/0000-0002-6754-7137}{\includegraphics[scale=0.06]{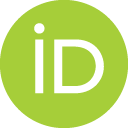}\hspace{1mm}Khalid S. Aloufi}\thanks{Use footnote for providing further
		information about author (webpage, alternative
		address)---\emph{not} for acknowledging funding agencies.} \\
	Department of Computer Engineering\\
	College of Computer Science and Engineering\\
	Taibah University\\
	Medina \\
	Saudi Arabia\\
	\texttt{koufi@taibahu.edu.sa} \\
	\And
	\href{https://orcid.org/0000-0002-7802-6628}{\includegraphics[scale=0.06]{orcid.png}\hspace{1mm}Abdulrahman A. Alsewari}\\
	Faculty of Computer Systems and Software Engineering\\
	Universiti Malaysia Pahang\\
	Kuantan 26300\\
	Malaysia\\
	\texttt{alsewari@ump.edu.my} \\
}

\begin{document}
\maketitle

\begin{abstract}
This research presents a methodology for trusting the provenance of data on the web.
The implication is that data does not change after publication and the source of the data is stable.
There are different data that should not change over time, such as published information in books and similar documents as well as news or events reported on the web.
If the data change after publication on the web, the web pages that reference the unstable data will lose points of interest or link to different resources.
With the current move to linked data and the semantic web, this is becoming a greater obstacle to be solved.
This research presents a methodology for establishing trusted information using an encoded reference of the data embedded in its URI, which creates a stable reference of the data and a method for ensuring its provenance stability.
After applying the methodology, the results showed that the methodology is highly applicable and has no overhead cost over the loading time.
The novel solution can be applied directly to any data portals or web content management systems.
\end{abstract}


\section{Introduction}
Currently, the web is the largest main source of information.
Other valuable sources of information, such as libraries and information banks, have most or part of their information on the web.
The information on the web can change over time.
However, there is no mechanism for tracing such changes in general.
Some web sites may have a mechanism to trace updates, such as showing versions of the web page or web page history.
Some web information should not change over time, such as scientific nanopublications, news and related media articles, literature, history and religious information.
The web is used to access such information using a uniform resource locator (URL), which is a web page uniform resource identifier (URI) for defining resources on the Internet.

There are differences between regular publishing and digital publishing.
For regular publishing, the information is printed in books or newspapers in a limited number of prints.
However, digital information is unlimited in the number of available copies and has no cost compared with regular printed information. Additionally, digital information can be secured in different ways.
Digital information has many more benefits over regular information distribution to a variety of readers anywhere. In fact, in history, the web is not comparable. However, it has raised different challenges as well.
For any available information, there is an author, a publisher and a reader.
For digital information, there are more parties, such as more publishers and readers. In fact, websites replicate information as a practice on today’s web.

For publishing, there are more differences between digital and regular methods.
When a regular publication is chosen by the author, the publisher is determined first and then the publication process begins.
Also, the information can be published in both ways, regular and digital, which is currently common these as of the writing of this article.

Additionally, digital publishing has a unique feature: the amount of information to be published. While regular publishing is usually for larger amounts of information, such as books, or as small as one column in a newspaper, it can be one word in digital publishing.
Cost is not comparable since digital publishing costs nothing compared to regular publishing.
Usually, digital publishing is open for everyone, with and without authority, depending on the publisher, such as social media, blogs or other news websites.

For digital publishing, when an author has information to publish, the author selects a trustworthy publisher to host the information. The audience is usually expected to go to such trusted hosts to find data about a topic. For example, a website allows users to publish their data and then can upload an expanded version of the data. The user-author can be in contact with his audience through this known website.

In digital publishing, the information published by an author can be as small as one word and as large as an unlimited number of words.
Additionally, machines can publish data, especially with the integration of the IoT in daily life.
Digital information can also be text as well as video or audio. In digital publishing over the web, every piece of information is considered a resource; metadata are used to describe the data, and every piece of information can be referred to using the web’s URI system. The web is making data publishing very flexible and affordable for humans and machines as well.
This research provides a methodology for obtaining trust in data publishing over the web.
One of the methodologies uses trusty URIs, which are mainly applied to nanopublications; however, the methodology can be developed for other web resources.

When the client wants to read a web page about a topic, a search engine is usually used to find related web pages.
The client selects one web page. At this point, the hosting web server processes the client request and returns a response to the client.
To search for a trusted resource, however, the user wants to make sure the information is original, trusted and verifiable, which current search engines have no mechanism for.
Additionally, using a validating authority, the client can validate a resource.
Therefore, the following system is presented for the publication of trusted resources for validation by clients of interest.

\section{Related Work}
Different research has been conducted on the Quran, which helps build advanced Quran applications. Additionally, the methodologies learned in these references can be applied to many different applications related to text resources as well as Arabic resources in general \cite{DBLP:journals/corr/abs-1801-03627}.
There has been considerable interest and research effort for building trusted methodologies for resources on the Internet, whereas some of the research has been mainly for Quran text.
Aimad Hakkoum and Said Raghay built a question-answering system based on the Quran with consistent meaning of verses \cite{Hakkoum2016}. They created a web interface of the developed ontology in the OWL web ontology language and the simple protocol and RDF query language (SPARQL)
\cite{Valle2011} \cite{knuth:1984}.
Ahmed SharafEldin and Shaimaa Salah Abbas developed a methodology for extracting knowledge from the Quran by developing an ontology of the Quran concept that can be used for any other application\cite{SharafEldin}.

The Quranic Speech Database for Arabic Speakers (QSDAS) has been developed to help natural language processing applications in Arabic\cite{Harrag}.
Raja Yusof and others have developed an efficient Arabic word stemmer tested on the 30th part of the Quran. Different Arabic word patterns are known to be extracted from the same roots \cite{RajaYusof}.
Karima Meftouh and others developed a methodology using a statistical method for analyzing Arabic text from newspapers \cite{MeftouhKarima}.
Ben Fraj Trabelsi and others have developed a method for parsing Arabic sentences based on a machine learning approach with a high success rate of accuracy\cite{benfrajtrabelsi}.
Hend Al-Khalifa and Amani A. Al-Ajlan developed a system for Arabic readability using machine learning techniques\cite{Al-Khalifa}.
Alghamdi, Mansour and Muzaffar, Zeeshan developed a methodology for diacritization of Arabic words based on quad-gram probabilities\cite{Alghamdi}.
Alghamdi, Mansour and Alotaibi, Yousef developed a recognition system for Arabic using hidden Markov models (HMMs) \cite{Alotaibi}. They used the Saudi Accented Arabic Voice Bank (SAAVB) as an Arabic speech corpus.
Droua-Hamdani and others developed an Arabic speech recognition system \cite{Droua-Hamdani}.

One of the studies analyzed Arabic text to extract questions from the text\cite{Al-Khatib}.
Segmenting Arabic text based on lexical analysis is essential for information extraction and summarizing \cite{Hamdi-Cherif}.
Al-yahya and others developed an ontology for the Quran based on the time nouns of verses\cite{yahya}.
Aqil Azmi and Nawaf bin Badia developed an e-narrator system for analyzing Hadith content\cite{Azmi}. Such approaches and studies can open a wide variety of applications over different Arabic structured resources, such as the Hadith and Quran. Some of the studies have considered enhancing the security of Quran text in terms of validation, such as the study performed by Mostafa G. \cite{7277220}. Hassan Abubakar and Suhaidi Hassan presented a methodology for using blockchains to provide digital trust in text from the Quran \cite{Abubakar}.

The proposed system used enhanced blockchain technology with different processes and components. For instance, the system contains Whisper as a network protocol, Swarm as a distributed storage, the Ethereum Virtual Machine (EVM)-Ethereum as a decentralized virtual machine and the Mist Browser as the system browser.
A. Boukabou and M. Khelifa presented a methodology for securely transmitting Quran text \cite{7277218}. They used chaotic oscillations as a method for encoding text during transmission.
M. Khan, Z. Siddiqui and O Tayan presented a digital certification methodology for Quran text. The user can easily verify the Quranic text online \cite{7889229}. Additionally, the same author noted that users are not satisfied with the validity of Quran text in applications \cite {Zakariah2017}. Additionally, they mentioned that the security and validation of Quran text are either performed by watermarking techniques or following the encoding of the text using different methodologies.
E. Nada, A. Ahmed and M. Abd-Allah presented an online e-learning system for the Quran \cite{7277222}. Additionally, they presented a questionnaire about online learning for the Quran. The study shows that most users are interested in learning the Quran online, which leads to the requirement of having secure and trusted online sources for the Quran text.
F. Kurniawan, M. Khalil, M. Khan, and Y. Alginahi presented a methodology for authenticating Quran images\cite{6597705}. The novel methodology presented can detect any change in the Quran image and the location of the change using fragile watermarking.

Y. Alginahi, M. Kabir and O. Tayan developed an approach for using watermarking to detect Arabic text documents \cite{6718288}. The methodology proposed is based on watermarking Kashida techniques to protect the authenticity and originality of the document.
In the paper titled "Islamic Knowledge Ontology Creation", the authors, S. Saad, N. Salim and H. Zainal, generated structured knowledge from the Quran using a methodology by building an ontology from Quran text and obtaining knowledge from it \cite{5402635}. The methodology used is a simulation based on the
combination of natural language processing
techniques (NLP), text
mining techniques and information extraction (IE). The integration of such studies with validation can be used to authenticate text from knowledge extraction.

B. Abuhaija, A. Awadelkarim, N. Shilbayeh, and M. Alwakeel applied a secure model mainly applied on web sites with Islamic content \cite{7351817}. The model guarantees trusted content when all parties join in the model. They presented the formal specification and the detection process of the watermark in a document enhanced by discrete wavelet transform (DWT) and contourlet transform (CT). The model provides secure content with trusted checks and validity that can be applied on websites under management since the web is open to the public and difficult to manage with human checks due to the large amount of information on the Internet today.

M. Majdalawieh, F. Marir and I. Tiemsani presented a study on modeling business processes for specific finance and management purposes based on the guidelines mentioned in the Quran and the Hadith \cite{8328549}. The paper presented an algorithm for gathering information about specific topics and actions to present Islamic solutions for businesses. Such studies show how important the validity of the resources is.
One of the methods used to verify content is by using trusty uniform resource identifiers (URIs), represented by Kuhn \cite{Kuhn} Kuhn\cite{DBLPjournalscorrKuhn15b}\cite{7079484}. This method is detailed in the next section, which will be used to develop a model for providing a methodology to provide a trusted resource of the Quran over a decentralized system.
Different web resources are represented in nanopublication formats, which makes it useful to have trusted sources using validating methodologies such as trusty URIs. A database of gene-disease associations (DisGeNET) has a gene dataset represented as nanopublication \cite{DisGeNET}.

\begin{figure}[h]
\centering
\begin{mdframed}
\begin{tikzpicture}[>=stealth',shorten >=1pt,node distance=2.8cm, semithick,scale=1.1,transform shape]
\node[state] (A)   at (0,0)  {Hash};
\node[state] (B)   at (4,0)  {Resource};
\node[state] (01)  at (2,-3) {Resource URI};
\node[state] (04)  at (6,-3) {Resource URI}; 
\path[->,auto] (A) edge [bend left]  node {Verifiable} (B)
      (B) edge [bend left]  node {Immutable} (A);
\draw [->,shorten >=0pt,dashed] (B.south) --node[pos=.7,fill=white,inner sep=1pt]{Permanent} (04);
\draw [->,shorten >=0pt,dashed] (B.south) --node[fill=white,inner sep=1pt]{Permanent} (01);
\draw [dashed] (01) --node[fill=white,inner sep=1pt]{} (04);
\end{tikzpicture}
\end{mdframed}
\caption{Trusty URI } 
\label{fig:Trusty URI}
\end{figure}
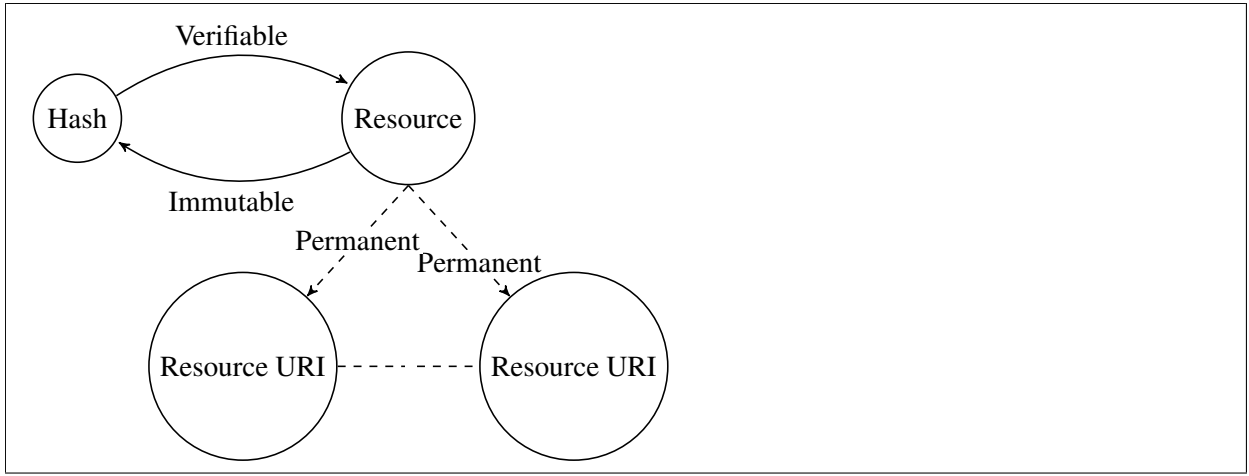

\section{Trusty URIs}
In this section, trusty URI, which is an authentication method for web resources, is detailed.
For example, when a web page is created, it is assigned a URL web address.
When the dynamic web page content changes, its URI is usually the same.
Generally, information is presented in different document structures, such as nanopublications, regular resource description framework (RDF) graphs, web page scripts or any digital artifact at the byte level, such as text or image files.
However, in some cases, the web page content should not change, such as the case of scientific publications, news reports, static information sources, published books, articles or facts.
This content should be updated in another web page so interested users can locate the new version of the information and go back to an earlier version of the information if there was any, which builds information trust.
Additionally, web resources are linked with other resources on the web. When these resources are verified and trusted, they are used to validate other linked resources, creating a network of trusted content, as will be shown.

To add trust to a web resource, trusty URI is used.
In trusty URI, the URI contains an added cryptographic digest hash value in Base64 notation representing the resource content.
The hash value is a short sequence of bytes or bits computed for the resource.
Any change in the resource results in a totally new hash value.
As of the state-of-the-art, although infinite different inputs can result in the same hash value, it is impossible to obtain the resource from the hash value.

Therefore, when the resource content changes, the hash value no longer represents the resource.
Someone can easily change the content and update the hash value as well.
However, when the content is distributed over different information sources, such as search engines or web archives, it is possible to check the validity of the resource content, and any change will be noticed.
Furthermore, the content URI is known and cannot change since it is supposed to be linked from other different resources on the web.
While the methodology increases the overhead process of generating trusted information, it is compatible with the web standards of open and distributed information resources.

As a result, trusty URI makes web resources verifiable, immutable, and permanent \cite{Kuhn}.
As mentioned earlier, the resource is verifiable because the hash value can be used to check the validity of the resource.
Consequently, the resource is immutable since any change in the resource will change the hash value.
The URI is permanent as well because it can be retrieved from other linked web resources, search engines or distributed web archives. Figure \ref{fig:Trusty URI} shows the three features. The hash value is used to verify the resource, the resource is immutable with the original hash and the resource is permanent since it has copies somewhere on the web.
The client requests a resource URI that contains a hash value.
Then, the server responds with the requested resource.
After that, the client can verify the resource at a trusted server.
These are the general steps for retrieving trusted content over the web, which are used in the proposed model of a trusty web resource publication system.

\subsection{Verifiable}

For the verifiable step, the author uses the publisher system to upload or generate content where the publisher task is to publish the resource with generated trusty URI based on the information provided by the author. Additionally, the resource should include the publication date and links to a minimum trusted URI. If the resource was the first from the author, then the system of the publisher links it to the main publisher page. This creates a chain of resources, which could help validate a complete tree of resources.
Therefore, the resource now has a trusty URI that can be used to verify the resource. 
\subsection{Immutable}

For the immutable step, when a client wants to retrieve a resource, a trusty search engine returns the trusty resources. The trusty search engine task  retrieves only trusted resources. The crawler of such search engines stores only trusted web pages after validation from step one. Additionally, distributed servers of trusted content or resources can be used to respond to reader requests of trusted resources only.

\subsection{Permanent}
For the permanent step, the user retrieves resources from a trusted source. Either the user or the source can perform the task; if the user validates the resource, then the user must search for the resource and validate it against the hash value found in the URI. If the source, such as a search engine, performs the task, then the user is free from the task unless is it performed optionally. 

\subsection{Trusty Resources}

To have a trusty resource, there are different components that need to work together in one system. Otherwise, the user has different options for performing the task, as described in a later section.
Initially, the author has a resource to share using a publisher system, which publishes resources with a trusty URI.
A distributed network of servers is used to host the new content to share the same copy of the newly generated resource.
Then, a client requests a resource from a server.
The user can then determine whether the resource is considered a valid trusted resource  using a validating entity, which can be achieved by contacting an entity from one of the servers from the distributed network of servers.

The information created as a digital artifact is represented as a resource published on the web with a web address as an original artifact and represented by its URI. It should have a publication date as well as a link to another already trusted resource.
As mentioned earlier, the resource is trusted if it has the three features verifiable, immutable and permanent resource.
Verifiable means the resource has a URI with a hash value.
Immutable resources can be validated by the URI.
Permanent resource copies are distributed over the Internet.
Each of the characteristics requires a sequence of tasks to be accomplished.

\subsection{Searching for Trusted Resources}
Initially, the user is looking for a trusted resource (A).
The user uses a search engine (S).
The user finds a resource that claims to be a trusted resource (A).
(A) is hosted by the source or host (H).
(H) allows its resources, such as (A), to have links to other trusted resources only.
Therefore, any URI from (A) is trusted.
If the trusted resource (A) is found in trusted source (H), a trusted search engine (S), then the job is completed for (A).

The search engine (S) only returns a trusted resource (A) from trusted hosts, such as (H), which is verifiable, immutable and permanent.
Otherwise, if the user finds a resource from an untrusted source (S) or (H), the user has to validate the resource (A) with a trusted source (S) or (H).
As a result, to complete the task of trusting a resource, the trusted source (S) is an essential part of the system. Through the trusted source (S), the user can trust the host (H) and the resources (A, B, C, ...).
As mentioned earlier, the resource is trusted if it is verifiable, immutable and a permanent resource.

Therefore, when the user finds trusted resource (A) that has another trusty URI of another trusted resource (B, C, ...), the user can trust any of the trusted links of (B, C, ...).
The user now knows the clicked link should lead to a trusted URI of the resource (B, C, ...). The user knows that the (B) URI is valid since resource (A) is valid.
However, the user does not know if the content of the resource (B) is verifiable, immutable and a permanent resource.
Therefore, the user must validate the linked resource (B).
If (B) is requested from a trusted host (H1), the user can consider (B) as a trusted resource.
When resource (B) is hosted in another untrusted host (H2), then the user has to verify (B) in a trusted server using the distributed network of servers.

Alternatively, any trusted host can provide a trusted service and be part of the distributed network of servers. The first host (H1) of resource (A) can be used to confirm that host (H2) is a trusted host; therefore, any linked URI is trusted, and hence, (B) is a trusted resource.
This feature can be added, and the trusted source that provided the resource (A) URI can help in validating any linked resource, such as resource (B).
Otherwise, resource (B) is not trusted.

Additionally, if the source (S) returns only trusted URIs, then the uncertainty of this step is solved.
Therefore, source (S) only allows trusty URIs in the resource, making it easier for the reader to follow up a trusted resource only by clicking trusted URIs.
This eliminates the issue mentioned in the step of validating a source.
This mechanism helps navigate from a trusted source to a trusted source, making the step easier for the user and hiding the complexity.
It is easier to build such trusted content.

\subsection{Restricted Publication Time}\label{sec:time}
The trusted resource should be published only once.
When published it should not change or update.
When there is an update, it should be published as a new resource.
Therefore, when the user navigates to resource (A) and finds a new link to other resources (B, C, D, ...), the user has no means for determining whether these URIs are valid. However, when the source (S) only returns trusted resources with trusty URIs to other resources (B, C, D, ...), the user knows that (A) is a trusted resource and that any of the linked resources (B, C, D, ...) is trusted as well.

However, (S) does not know if the resource has changed.
One solution to this issue is validating the resource either by a plugin at the client or by a distributed network of servers.

In the first option, by using a plugin in the client browser, the user knows that the URL is a trusty URI; otherwise, the search engine will not return the resource. The user knows that the resource could have been changed later and the URL to other resources added and not checked by the search engine because the search engine does not update its record every second. Therefore, the solution to this issue is a plugin at the client browser to double-check the source trust.

Therefore, the user knows that the resource is obtained with a trusty URI and can check the resource and confirm that the resource is verifiable, immutable and a permanent resource.

The second option, as an alternative to using a plugin, is that the client can validate the resource using a validating server from a network of distributed servers that help the client and search engines check resources.

\subsection{Trusting sources and hosts}
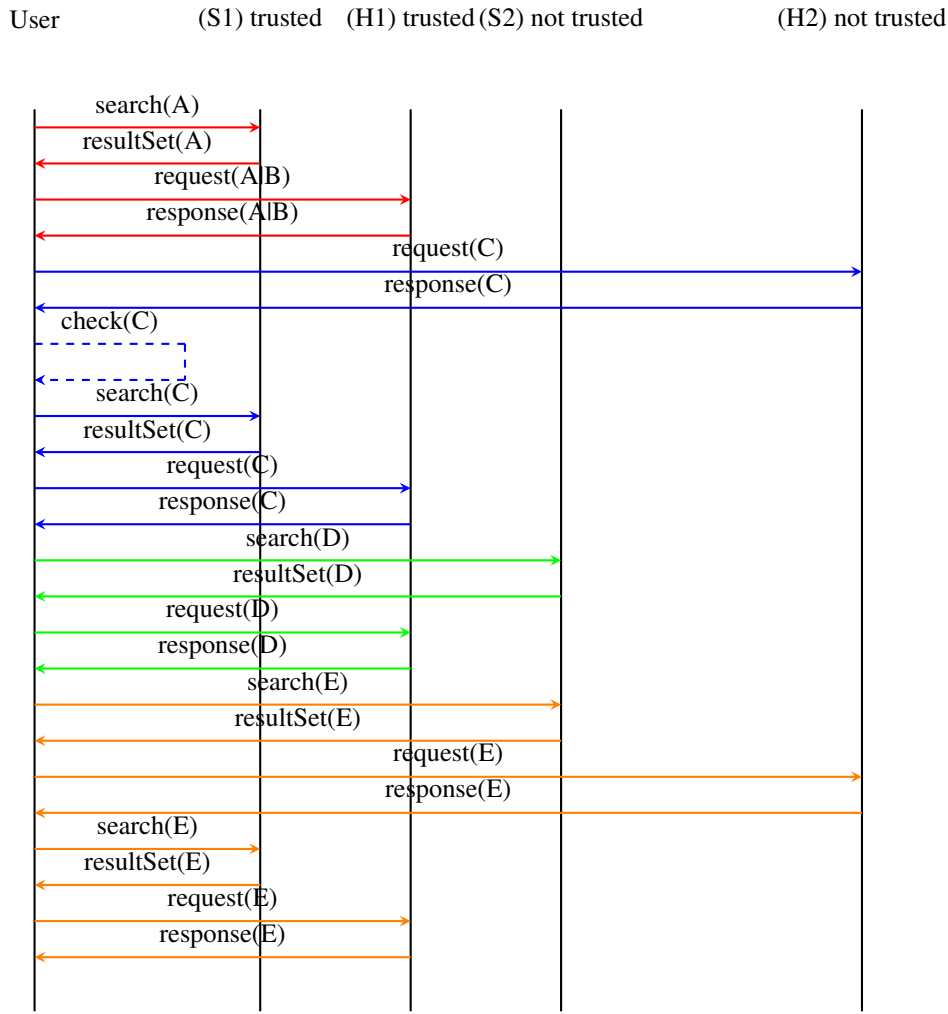
\begin{figure}[h]
\centering
\begin{tikzpicture} 
\coordinate (c) at (-1,0);
\coordinate (d) at (-1,12);
\coordinate (cc) at (1,0);
\coordinate (dd) at (1,12);

\coordinate (e) at (2,0);
\coordinate (f) at (2,12);
\coordinate (g) at (4,0);
\coordinate (h) at (4,12);
\coordinate (i) at (6,0);
\coordinate (j) at (6,12);
\coordinate (k) at (10,0);
\coordinate (l) at (10,12);

\draw[style={draw=black,thick,}] (c) -- (d) node[pos=1.1]{User};
\draw[style={draw=black,thick,}] (e) -- (f) node[pos=1.1]{(S1) trusted} ;
\draw[style={draw=black,thick,}] (g) -- (h) node[pos=1.1]{(H1) trusted};
\draw[style={draw=black,thick,}] (i) -- (j) node[pos=1.1]{(S2) not trusted};
\draw[style={draw=black,thick,}] (k) -- (l) node[pos=1.1]{(H2) not trusted};

\draw[-stealth,thick,style={draw=red}] 
($(c)!0.98!(d)$) -- node[above,midway]{search(A)}($(e)!0.98!(f)$);
\draw[-stealth,thick,style={draw=red}] 
($(e)!0.94!(f)$) -- node[above,midway]{resultSet(A)}($(c)!0.94!(d)$);
\draw[-stealth,thick,style={draw=red}]
($(c)!0.90!(d)$) -- node[above,midway]{request(A|B)}($(g)!0.90!(h)$);
\draw[-stealth,thick,style={draw=red}]
($(g)!0.86!(h)$) -- node[above,midway]{response(A|B)}($(c)!0.86!(d)$);

\draw[-stealth,thick,style={draw=blue}]
($(c)!0.82!(d)$) -- node[above,midway]{request(C)}($(k)!0.82!(l)$);
\draw[-stealth,thick,style={draw=blue}]
($(k)!0.78!(l)$) -- node[above,midway]{response(C)}($(c)!0.78!(d)$);

\draw[dashed,thick,style={draw=blue}]
($(c)!0.74!(d)$) -- node[above,midway]{check(C)}($(cc)!0.74!(dd)$);
\draw[dashed,thick,style={draw=blue}]
($(cc)!0.74!(dd)$) -- node[above,midway]{}($(cc)!0.70!(dd)$);
\draw[dashed,-stealth,thick,style={draw=blue}]
($(cc)!0.70!(dd)$) -- node[above,midway]{}($(c)!0.70!(d)$);

\draw[-stealth,thick,style={draw=blue}]
($(c)!0.66!(d)$) -- node[above,midway]{search(C)}($(e)!0.66!(f)$);
\draw[-stealth,thick,style={draw=blue}]
($(e)!0.62!(f)$) -- node[above,midway]{resultSet(C)}($(c)!0.62!(d)$);

\draw[-stealth,thick,style={draw=blue}]
($(c)!0.58!(d)$) -- node[above,midway]{request(C)}($(g)!0.58!(h)$);
\draw[-stealth,thick,style={draw=blue}]
($(g)!0.54!(h)$) -- node[above,midway]{response(C)}($(c)!0.54!(d)$);

\draw[-stealth,thick,style={draw=green}]
($(c)!0.50!(d)$) -- node[above,midway]{search(D)}($(i)!0.50!(j)$);
\draw[-stealth,thick,style={draw=green}]
($(i)!0.46!(j)$) -- node[above,midway]{resultSet(D)}($(c)!0.46!(d)$);
\draw[-stealth,thick,style={draw=green}]
($(c)!0.42!(d)$) -- node[above,midway]{request(D)}($(g)!0.42!(h)$);
\draw[-stealth,thick,style={draw=green}]
($(g)!0.38!(h)$) -- node[above,midway]{response(D)}($(c)!0.38!(d)$);

\draw[-stealth,thick,style={draw=orange}]
($(c)!0.34!(d)$) -- node[above,midway]{search(E)}($(i)!0.34!(j)$);
\draw[-stealth,thick,style={draw=orange}]
($(i)!0.30!(j)$) -- node[above,midway]{resultSet(E)}($(c)!0.30!(d)$);

\draw[-stealth,thick,style={draw=orange}]
($(c)!0.26!(d)$) -- node[above,midway]{request(E)}($(k)!0.26!(l)$);
\draw[-stealth,thick,style={draw=orange}]
($(k)!0.22!(l)$) -- node[above,midway]{response(E)}($(c)!0.22!(d)$);

\draw[-stealth,thick,style={draw=orange}]
($(c)!0.18!(d)$) -- node[above,midway]{search(E)}($(e)!0.18!(f)$);

\draw[-stealth,thick,style={draw=orange}]
($(e)!0.14!(f)$) -- node[above,midway]{resultSet(E)}($(c)!0.14!(d)$);
 \draw[-stealth,thick,style={draw=orange}]
($(c)!0.10!(d)$) -- node[above,midway]{request(E)}($(g)!0.10!(h)$);
\draw[-stealth,thick,style={draw=orange}]
($(g)!0.06!(h)$) -- node[above,midway]{response(E)}($(c)!0.06!(d)$);

\end{tikzpicture}
\caption{Trusted source and host}
\label{fig:Trusted source and host}
\end{figure}

The user either trusts the source, which is assumed to be a search engine, (S1), and the source host (H1), which is assumed to be a web server.
Additionally, the user may trust (S1) but not trust (H2), or not trust (S2) but trust (H1) or, as a last possible situation, not trust both (S2) and (H2). Figure \ref{fig:Trusted source and host} shows the four conditions.

The first case is when the user trusts the search engine (S1) and the source host (H1); the situation is considered ideal because the user in this situation can trust the retrieved resource (A) and its internally linked resources, such as (B), as shown in figure \ref{fig:Trusted source and host} with red lines.
The user trusts resource (B) because it is retrieved from the trusted host (H1) and linked from resource (A), which is retrieved from the trusted host (H1).

In the second case, when the user trusts the source (S1) but does not trust the host (H2), the user can accept the retrieved resource (C) and its linked URIs but will need more overhead processing to trust the resources (C), as shown in figure \ref{fig:Trusted source and host} with the blue line.
The user knows that the URI of resource (C) is correct. However, as mentioned, the user does not trust (H2). Therefore, the user does not trust the resource content and, therefore, needs to check the resource (C). The user will have to check that the content of resource (C) matches its URI, as shown in figure \ref{fig:Trusted source and host} with the dashed blue line. Alternatively, the user needs a trusted source (S1) and trusted host (H1) to verify (C), as shown in figure \ref{fig:Trusted source and host} with the blue line.
If (C) is trusted, then all its content is trusted, and the process continues as in the first case or the second case.

In the third case, when the user does not trust the source (S2) but does trust the host (H1), then the user can trust the retrieved resources (D) from the host (H1), as shown in figure \ref{fig:Trusted source and host} with the green line. After (D) is trusted, all its content is trusted, and the process continues, as mentioned in the first case or the second case.

When the user does not trust both the source (S2) and the host (H2), which is the case for today's web content, the user must use case one, which requires a trusted (S1) and a trusted host (H1) to verify the resource (E), as shown in figure \ref{fig:Trusted source and host} with the orange line. Otherwise, the user will not trust resource (E). After (E) is trusted, all its content is trusted, and the process continues as mentioned in the first case or the second case.

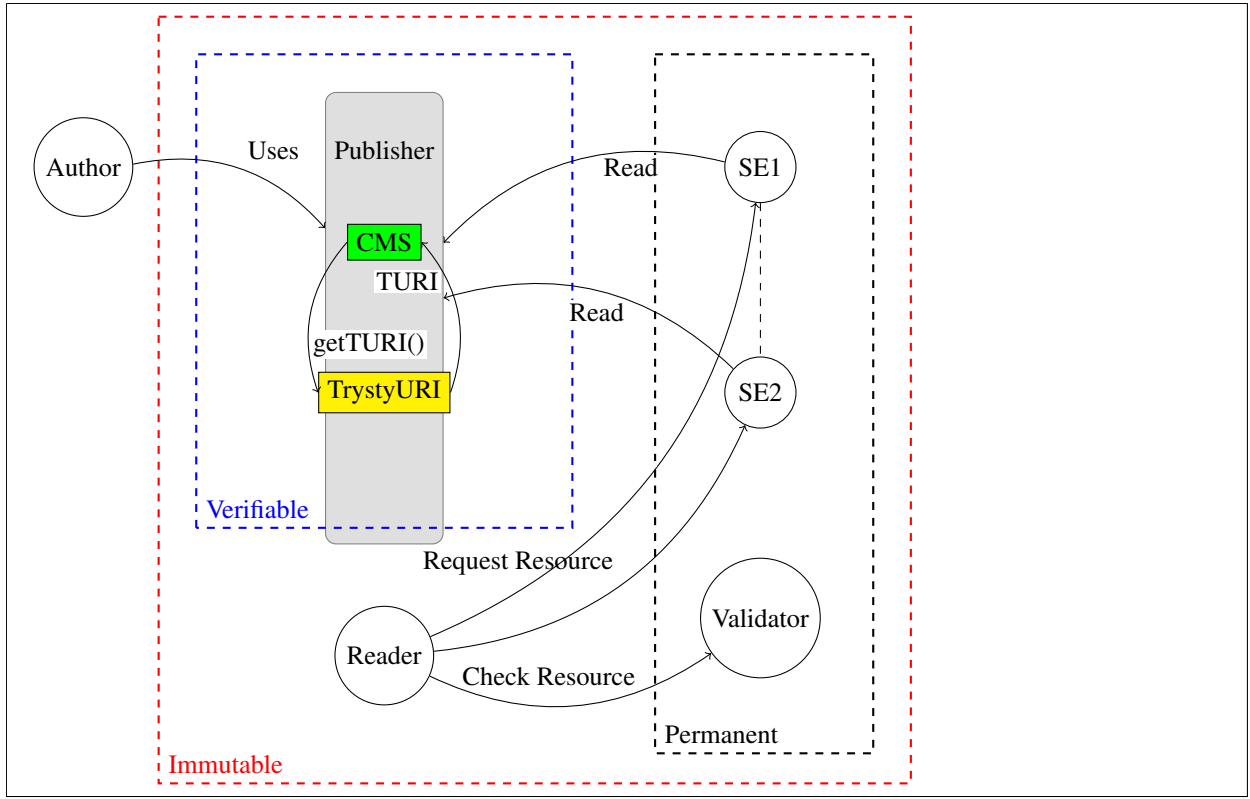
\begin{figure}[h!]
\centering
\begin{mdframed}
\begin{tikzpicture}
\node[state] (AUTHOR)   at (-2,-1)  {Author};
\node[basic box = gray,  anchor = north] (PUBLISHER) at (2,-0)    
{\\
\\
Publisher
\\
\\
\\
\\

\\
\\
\\
\\
\\

\\
\\
\\
\\};
\node[draw, fill = green](WS)  at (2,-2)   {CMS};
\node[draw, fill = yellow](Trysty1)   at (2,-4)   
    { Trysty  \\URI };
\path[->,auto] (WS.west) edge [bend right]  node[pos=.8,fill=white,inner sep=1pt] {getTURI()} (Trysty1.west);
    
 \path[->,auto] (Trysty1.east) edge [bend right]  node[pos=.8,fill=white,inner sep=1pt] {TURI} (WS.east);   
  
\node[state] (SE1)   at (7,-1)  {SE1};
\node[state] (SE2)   at (7,-4)  {SE2};
\node[state] (READER)   at (2,-7.5)  {Reader};

\node[state] (VALIDATOR)   at (7,-7)  {Validator};
\draw[black,thick,dashed] (5.6,.5) -- (8.5,.5) -- (8.5,-8.8) -- (5.6,-8.8)node[anchor=south west] {Permanent} -- (5.6,.5);
\draw[blue,thick,dashed]  (-.5,-5.8) -- (-.5,0.5) -- (4.5,0.5) -- (4.5,-5.8) -- (-.5,-5.8) node[anchor=south west] {Verifiable};
\draw[red,thick,dashed] (-1,1) -- (9,1) -- (9,-9.2) -- (-1,-9.2) node[anchor=south west] {Immutable} -- (-1,1) ;
\path[->,auto] (AUTHOR) edge [bend left]  node {Uses} (PUBLISHER);
\path[->,auto] (SE1) edge [bend right]  node[pos=.4,fill=white,inner sep=1pt] {Read} (PUBLISHER);
\path[->,auto] (SE2) edge [bend right]  node[pos=.4,fill=white,inner sep=1pt] {Read} (PUBLISHER);
\path[->,auto] (READER) edge [bend right]  node[pos=.5,fill=white,inner sep=1pt] {} (SE1);
\path[->,auto] (READER) edge [bend right]  node {Request Resource} (SE2);
\path[->,auto] (READER) edge [bend right]  node[pos=.1,fill=white,inner sep=1pt] {Check Resource} (VALIDATOR);
      


\draw [dashed] (SE1) --node[fill=white,inner sep=1pt]{} (SE2);

\end{tikzpicture}

\end{mdframed}
\caption{Trusty URI function over entities} 
\label{fig:TrustyURI Function over entities}
\end{figure}

\section{Proposed System}

There are different methodologies for building a trusty URI system to check the trust of a resource, as shown in the earlier section.
Different entities will use the trusty URI system: the author, the reader, the publisher and the trusty server, which could be a distributed trusty server network.
Figure \ref{fig:TrustyURI Function over entities} shows the systematic functions of a trusty system to have a trusty URI and trusty resources.
The trusty system is composed of trusty URIs, trusty resources and trusty sources.
Trusty URIs are the URIs with hash values that represent its resource content.
A trusty resource is the
If these three components are available, then we have trusty web content.
In case we must add a validator when there is one missing component of the trusty system,
the system has a small overhead over a regular system, which is the cost paid for having a validator only, which means the proposed system is following current web standards and built according to the regular information retrieval models on the web today. As mentioned earlier, it is very useful for specific kinds of content.
The figure shows the solution with the required components and the associated processes. Following the actions shown in the figure, the author has content to publish as trusted content for readers.
Therefore, the author uses a publisher system that has a content management system (CMS) with a trusty unit. The job of the trusty unit is to generate the hash value of the resource and include it in the URI.

Additionally, an HTML web page can have trusty URIs as well as regular URIs.
For example, for the HTML resource, as shown in figure \ref{fig:example resource}, the trusty resource URI is shown at the top.
Like HTML, linked data are referenced over the web and can be trusted or validated depending on the source.
Figure \ref{fig:example resource} shows an embedded trusty URI for another page.
Self-referencing, as in the case of some web resources, is not assumed in this work since it increases the complexity of the validation process.
Self-referencing can be applied as mentioned in the literature since some data standards have internal referencing such as nanopublication.
As shown in the figure, SE1 and SE2 are known trusted search engines that reference content from trusty resources only. After the reader identifies a trusty source using trusty search engines, the reader can be confident that the resources retrieved are from trusty sources.
When using regular search engines, the reader is responsible for checking the resource.
Generally, retrieving a resource from a trusty source means that the trusty resource has a trusty URL as well.

In case of having a trusty URI without trusting the content, there is the option of using a validator as mentioned in section \ref{sec:time} to check the URI; otherwise, if the reader is reading the content from a trusty publisher, then this step is not required.
Additionally, if the trusty resource is retrieved from a trusty source, then the URIs associated implicitly inside the trusty resource are trusty URIs.
Following a trusty URI from a trusty resource, there are two possibilities. The first is that the resource is retrieved from the trusty source. In this case, the user has no requirement to validate the content. In the second case, the user retrieves content from a source not known as a trusty resource, such as the case when navigating the web from one website to another.
Then, the user has the option to validate it using a validator.
As seen from the different scenarios, the trusted source is the main step.
That is assumed because when a user finds a resource on the web, the only way to know if it is trusted is to find it from a trusty source.
In the other scenario, when finding a trusty URI but it is not known whether the source is trusty, then using the validator will make the determination.

\section{Results and Discussion}
The Quran is one of the holy books.
The Quran contains the Sura, and each Sura contains the Ayah (verses).
The Quran contains 114 Sura, 14,870 distinct words, 78,245 words in total, 6,236 Ayah, 3,178 distinct roots and 78,245 roots \cite{QuranDataset}.
The Quran is one of the resources that is permanent and immutable.
The Quran verses are used in web pages as part of an article or only a sequence of Ayah.
In addition to the different methodologies used to verify the contents of different web content mentioned earlier, Quran verses can be verified using the methodology presented in this paper.
In a web page that contains part of the Quran, part of or the whole page can be verified if needed.

This work tests the proposed system with the Quran as a resource in text format, such as HTML resources, as shown earlier in figure \ref{fig:example resource}, which is taken from the dataset \cite{QuranDataset}.

Therefore, the main task of verifying or publishing a trusted resource is generating the hash value. This is a challenging task since service time is critical for web applications. Therefore, different experiments and tests show the service time for different sizes of data. Additionally, the experiment shows that different methodologies have different service times for the same task.

The operating system used for the experiment is Windows 10 64 bit running on an Intel® Core™ i7-4770 Processor with 16 GB of RAM. The programming language used is Java, and the database is used to store the data over PostgreSQL as the database management system (DBMS). Then, a set of codes returns the result of the experiment to test the service time for generating the hash values for different resource sizes.

Figure \ref{fig:maxsurahayah} shows that the maximum time required to generate the hash for the text of a Surah is 254.460688 ms. For the text of an Ayah, the maximum time required to generate the hash is 27.444402 ms.
Figure \ref{fig:minsurahayah} shows that the minimum time required to generate the hash for the text of a Surah is .667440 ms. For the text of an Ayah, the maximum time required to generate the hash is .119239 ms.

Generating the hash value for Quran text that has already been read by the computer is 9.182648 ms, as shown in figure \ref{fig:text}, and the hash value generated is 5c79fc50b16917aeb6e153f51d1c92c1abbef2f43ea5d3a96cdb643617ee70f0.
Generating the hash value for the Quran text that requires reading first by the computer is 131.867402 ms, as shown in figure \ref{fig:text}, and the hash value generated is the same as earlier.

\begin{figure}[h]
\centering
\begin{tikzpicture}
\begin{axis}[
 bar width=0.05,
    ybar,
    tick label style={font=\small},
    tickpos=left,
        ylabel=ms,
xticklabels={hashfilechecksum}, 
    xtick={1},
         ymin=0,
    y tick label style={/pgf/number format/.cd,%
          scaled y ticks = false,
          set thousands separator={},
          fixed
    },
    ]
    \addplot +[bar shift=-.2cm, area legend] coordinates {(1,7782.190209) };

\end{axis}
\end{tikzpicture} 
 \caption{File using hashfilechecksum } 
\label{fig:hashfilechecksum}
\end{figure}
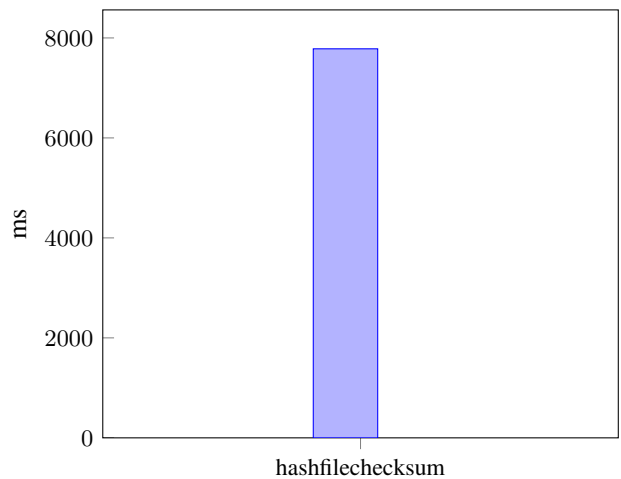

To hashing the doc format source file using the MessageDigest class and DigestInputStream class in Java, the time required to generate the hash value is 7782.190209 ms, as shown in figure \ref{fig:hashfilechecksum}.
The hash value generated is the same as that generated using the DigestUtils class.
However, the time required to generate the hash value is 56.928554 ms, as shown in figure \ref{fig:checksumSHA256}, which is .7\% of the time required using the MessageDigest class and DigestInputStream class.
It is one thing to use Java security MessageDigestSpi in terms of performance but using org.apache.commons.codec.digest DigestUtils is quite another.

\begin{figure}[h]
\centering
\begin{mdframed}
Example\_Page\_Title
\end{mdframed}
\begin{mdframed}
https://www.example.com/803dee27b5a9ddf866112eac2f8a
34e5bd83ca08015ec4acaaffeba25a378352
\end{mdframed}
\begin{mdframed}
\begin{lstlisting}[captionpos=b, caption=, basicstyle=\ttfamily, breaklines=true] 
<!DOCTYPE html>
<html>
<body>
<h1>My Trusty Resource</h1>
<p>My Trusty Resource.</p>
<p title="Trusty Resource">
<a href="https://www.example.com/22b81fd12c136d4cf67a37de941908d8
3eaf8e97571c4983f9308d30d52ad8f9">This is a link to a Trusty Resource</a>
</p>
</body>
</html>
\end{lstlisting} 
\end{mdframed}
\caption{Example of a trusty HTML resource } 
\label{fig:example resource}
\end{figure}

\begin{figure}[h]
\centering 
\begin{tikzpicture}
\begin{axis}[
bar width=0.6,
ybar,
tick label style={font=\small},
tickpos=left,
xticklabels={Surah, Ayah}, 
xtick={1,2},
ylabel=ms,
ymin=0,
legend entries={max,min},
legend pos=north east,
y tick label style={/pgf/number format/.cd,%
scaled y ticks = false,
set thousands separator={},
fixed
},
enlarge x limits={abs=1}
]
\addplot +[bar shift=-.2cm, area legend] coordinates {
(1,254.460688) 
(2,27.444402) };
\end{axis}
\end{tikzpicture} 
\caption{Maximum time for generating the hash file. } 
\label{fig:maxsurahayah}
\end{figure}

\begin{figure}[h]
\centering
\begin{tikzpicture}
\begin{axis}[
bar width=0.5,
ybar,
tick label style={font=\small},
tickpos=left,
xticklabels={Surah, Ayah}, 
xtick={1,2},
ylabel=ms,
ymin=0,
legend entries={min},
legend pos=north east,
y tick label style={/pgf/number format/.cd,%
scaled y ticks = false,
set thousands separator={},
fixed
},  
enlarge x limits={abs=1}
]
\addplot  +[bar shift=.2cm, area legend]coordinates {
(1,.667440) 
(2,.119239) };
\end{axis}
\end{tikzpicture} 
 \caption{Minimum time for generating the hash file. } 
\label{fig:minsurahayah}
\end{figure}

\begin{figure}[h]
\centering
\begin{tikzpicture}
\begin{axis}[
 bar width=0.2,
    ybar,
    tick label style={font=\small},
    tickpos=left,
        ylabel=ms,
xticklabels={ textalreadyread,textwithreading}, 
    xtick={5,6},
    ymin=0,
    y tick label style={/pgf/number format/.cd,%
          scaled y ticks = false,
          set thousands separator={},
          fixed
    },
    ]
    \addplot +[bar shift=-.2cm, area legend] coordinates { (5,9.182648)(6,131.867402) };

\end{axis}
\end{tikzpicture} 
 \caption{text} 
\label{fig:text}
\end{figure}

\begin{figure}[h]
\centering
\begin{tikzpicture}
\begin{axis}[
bar width=0.2,ybar,
tick label style={font=\small},
tickpos=left,ylabel=ms,
xticklabels={ checksumSHA256}, 
xtick={4,5,6},ymin=0,
y tick label style={/pgf/number format/.cd,%
scaled y ticks = false,
set thousands separator={},
fixed
},
]
\addplot +[bar shift=-.2cm, area legend] coordinates { 
(4,56.928554)   };

\end{axis}
\end{tikzpicture} 
 \caption{File using checksumSHA256} 
\label{fig:checksumSHA256}
\end{figure}
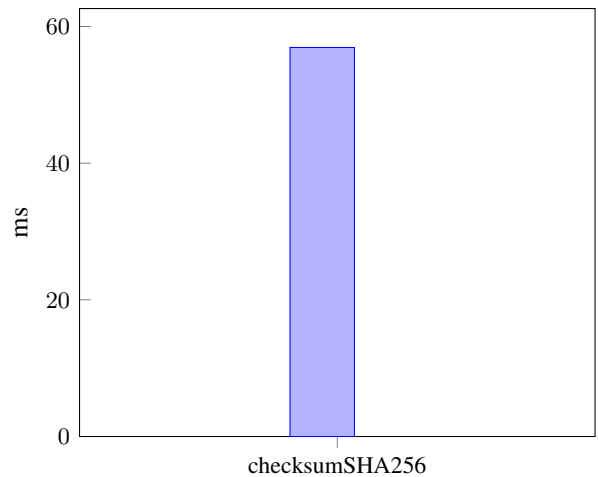

\section{Conclusion}

Resource validation on the web is one of the key challenges for trusting data found on the web.
There are different methodologies used to trust web data; however, there is no guarantee that the data version has been maintained.

The client can validate a resource using a trust certificate issued by a validating entity for a specific domain with the option of an external verification URL.

Having a trusty URL, which means that the URL has been checked, then any information on that web page is also trusted. Other URLs on this page are also trusty URIs, however, these contents also need to be checked.
Therefore, when using a trusty URIs, the contents of this URI need to be checked. Once checked, all the information in this URI also has valid URIs.
However, when visiting a web site and its content needs to be validated, there is a method to check it, such as a trusty server that can check the URI for and determine whether it is trusty or not for other URIs in the page.
Currently, the user accesses a page and determines whether it trusts the URI, which is cumbersome.
This research presents a methodology for trusty data over trusty URIs.

Providing a different dynamic methodology for web sites to validate content helps increase the trust of the provenance of the information.
Any change in the text will change the URI.
In this work, a test was performed using regular text that could represent regular HTML web pages. However, for the provenance verification level of linked data represented using RDF, the SPARQL endpoint has a web-based query, where the page obtains the URL of the resource and validates against its dataset. Following the same model represented in this study, however, with a SPARQL endpoint automatic machine-to-machine validation of the content becomes an obvious task.
Linked data makes this easier using SPARQL endpoints for published information in RDF standards,
which include RDF in html web pages in a verifiable way by using hash codes for every resource in RDF or HTML.
The system can be developed for distributed verification, where the resource is validated against more than one server.

As mentioned, the resource has a unique URI generated by the authority as the publisher of the resource.
After publishing, there is no authority.
In this work, the system is tested using the Quran, where each test represents an Ayah or Sura. In the future, the system will be tested on images.
In the future, an associated protocol can be developed that can help in Industry 4.0 applications as well as scientific publications.
In conclusion, this work presents a method for trusted resources over the web for sensitive information. The web does not support trusted content in its current standards. Therefore, different research and applications are necessary to fill this gap and create an environment for specific content, and the web is a place for verifiable, immutable and permanent content.

\bibliographystyle{unsrt}
\bibliography{refer}

\end{document}